# Botnet in the Browser: Understanding Threats Caused by Malicious Browser Extensions

Raffaello Perrotta, and Feng Hao

**Abstract**—Browser extensions have been established as a common feature present in modern browsers. However, some extension systems risk exposing APIs which are too permissive and cohesive with the browser's internal structure, thus leaving a hole for malicious developers to exploit security critical functionality within the browser itself. In this paper, we raise the awareness of the threats caused by browser extensions by presenting a botnet framework based on malicious extensions installed in the user's browser, and an exhaustive range of attacks that can be launched in this framework. We systematically categorize, describe and implement these attacks against Chrome, Firefox and Firefox-for-Android, and verify experiments on Windows, Linux and Android systems. To the best of our knowledge, this paper presents to date the most comprehensive analysis about the threats of botnet in modern browsers due to the over-privileged capabilities possessed by browser extensions. We also discuss countermeasures to the identified problems.

**Index Terms**—Web Browser, Scripting Languages.

## 1 Introduction

Attacks launched from within the browser have increased dramatically in recent years [1]. Though browsers have taken steps to restrict the installation of extensions to trusted sources, instances of malware being packaged into seemingly innocuous extensions are still regularly being reported. For example, a study conducted in 2014 by Kapravelos et al. discovered 130 compromised Chrome extensions [2]. In 2016, MalwareBytes identified a rogue extension [3] uploaded to Chrome's web store known as *iCalc*, which had bypassed Google's automated extension review audit and silently transmitted sensitive data to a rogue server. A recent study by researchers in Google highlights extensive efforts by criminals to distribute malicious extensions through the Chrome Web Store: nearly 10% of the Chrome extensions were identified as malicious between 2012 – 2015 [1].

Browser extensions are an attractive target for botnets, which are responsible for many large scale attacks on the Internet infrastructure today [4]. First, they have potentially a large installation base with hundreds of millions of users, especially when an adversary opts to distribute malicious extensions by first buying an existing popular extension and then adding malicious code in the update as reported in [1]. Second, a browser extension possesses many over-privileged capabilities in accessing sensitive user data. Such capabilities can be easily abused as we will explain in this paper. Finally, it is relatively easy to trick a user into installing an innocuous-looking extension that performs subtle malicious activities in the background [5]. Once installed, an extension forms an integral part of the browser and is outside the control of anti-virus software that is installed on the user's computer.

One early study about botnets in the browser was due to Liu et al. in 2011 [4]. The authors proposed a botnet framework based on browser extensions. Their system relied on exploiting the extension update mechanism, which browsers provide, to issue commands in batch. The commands would then be processed by the installed extensions according to their designed format. Within their implementation, they include three attack examples: email spamming, DDoS and password sniffing. Their architecture relies on the extension checking for updates using the extension's own update mechanism, which is only polled during events such as browser startup. This means an attacker cannot target commands to specific users in the network. Although there are follow-up papers by Liu et al. in 2012 [6] and other researchers [7], [8], they are all limited in covering only a subset of botnet attacks.

In this paper, we aim to systematize the knowledge in this domain to raise the awareness of the threats present in modern browsers caused by malicious browser extensions. To this end, we present an extension-based botnet framework which allows fine-grained controls via a *phone-home* based model, and an exhaustive range of attacks that can be launched by malicious browser extensions. The attacks are systematically categorized, described and implemented on Chrome, Firefox and Firefox for Android, which correspond to 59.24%, 13.29% and 0.68% of the browser market share [9], respectively. All the attacks have been experimentally validated against Chrome 54 and Firefox 49 (on both desktop and mobile)–which were the versions most recently available during the study–and on Windows, Linux and Android systems. To the best of our knowledge, the results we present are the most comprehensive in the literature about the threats of botnet in modern browsers imposed by malicious browser extensions. Finally, countermeasures to the identified problems are also discussed in this paper.

## 2 Background

Modern browser extension systems are based around the JSE (JavaScript Engine) model [10], in which the browser extension is a small set of scripts, generally JavaScript, running as part of the browser process. Extensions have evolved over time to reflect the needs of their user base which has subsequently led to more privileged extension models being introduced. The majority of extension systems are free to read and manipulate the web page's Document Object Model (DOM). In addition, many common privileges include access to the user's browsing history, bookmarks and even lower level functionality such as arbitrary file access. The levels of privilege vary depending on the security model set out by the browser's implementation. Chrome utilizes an extension system that heavily makes use of the *isolated worlds* paradigm, which in layman's terms simply means each script runs within their own sandbox. Firefox on the other hand, including Firefox for Android, utilizes an extension model that has been subject to wide criticism, with Liverani and Freeman claiming security was *"almost inexistent"* [11].

*The authors are with the School of Computing Science, Newcastle University, UK. Email: {r.perrotta, feng.hao}@newcastle.ac.uk.* *The second author would like to acknowledge the support of ERC Starting Grant, No. 306994.*

### 2.1 Chrome Extensions

Chrome extensions comprise three distinct components: the manifest file, the event pages, and content scripts. Additional pages such as HTML or style sheets may also be included within the extension provided they are declared within the manifest. Chrome enforces within its extensions the *principle of least authority (POLA)*, a motif in the security community which entails allocating only the permissions necessary to complete a specific action. A manifest file must be declared to specify which permissions and URLs the extension is allowed to access. In theory, this fine granularity of permissions should deter users from installing extensions which request permissions outside its intended scope. However studies have shown that permission systems are not as effective as predicted [12], [13] as most users tend to trust applications which seem to come from reputable or popular sources. Some users are also not aware of potential security implications which are caused, or simply do not find taking the risk of installing permissive applications to be problematic.

Event pages act as the intermediate layer between the browser and the content scripts. Event pages run as a separate process within Google Chrome, and access browser contexts via APIs, which bridge the sensitive internal browser operations to a set of reduced calls that guarantee restriction in what the extensions themselves have the power to do. In the situation an event page wants to modify the page's content, communication may bilaterally occur between the event pages and content scripts; it is important to note that event pages do not have direct access to the DOM.

Content scripts are JavaScript files which are injected within a web page, and run in the same process space as the document that is rendered. They have the ability to read and manipulate the DOM without restrictions, providing the appropriate permission is adjudicated. Furthermore, content scripts may arbitrarily issue outbound cross-origin HTTP requests, without requesting additional permissions. This exposes a design decision that has some security implications to it, as we will explain later.

### 2.2 Firefox Extensions

Jetpack extensions were introduced by Mozilla in 2013 as a solution to the archaic XPCOM-based extension system used by early versions of Firefox. XPCOM, stands for *cross platform component object model* and provides access to lower level interfaces. Within XPCOM, the functionalities allowed include reading system files and utilizing system libraries. Jetpack was devised as a model which utilized the POLA within its modules to contain vulnerabilities. Jetpack extensions, in similar regards to Chrome, comprise a manifest file, at least one module (which mirrors event pages in Chrome) and content scripts. An interesting note is that the manifest file in Jetpack extensions specifies metadata rather than permissions. This is in contrast to Chrome which employs a detailed permission schema and displays it to the user at install time.

Modules serve as the interface between native browser APIs, and any content scripts that have been registered within the extension. Mozilla provides a set of core modules which can be re-used with the aim to provide a safer interface to low-level functions. These core modules can be used indiscriminately by an extension developer. It is expected that within the Jetpack implementation, a strong isolation policy between modules and access to the native XPCOM components should be in place; however, even as of recently, Add-on SDK still allows developers to access the majority of the XPCOM functions. Though, like Chrome, modules do not have access to the DOM of a page, a similar message passing system allows communication between the module APIs and a content script. Content scripts in Firefox perform a role identical to that described in Chrome's model.

#### 2.2.1 *Firefox for Android*

Firefox for Android, also known as the Fennec project, was introduced in 2011. It is essentially a port of the main Firefox adapted to run on the Android platform. The code pertinent to our research, within the Add-on SDK, is nearly a one to one mapping between standard Firefox and its Fennec counterpart. Due to the differences between the security policies of the Android operating system compared to a PC (say Windows) operating system, we do observe differences in the impacts of some of our attacks. A similar observation can be made due to the introduction of more sensor devices, by default, on a standard smartphone. We will discuss in more detail in Section 5 the impact of attacks in different operating systems.

## 3 Threat Model

In our threat model, we assume a malicious extension has been installed on a target's system. There are many methods that an adversary can use to persuade a user to install a malicious browser extension, including disguising malicious extensions as legitimate browser extensions, using Trojans to install malicious extensions, and missing plugin attacks [5]. Another method available to attackers is to purchase a popular extension from its creator and then add malicious code to it. This technique has been observed to be used by adware producers [1]. Once a malicious extension is installed, we assume it has been granted all the privileges it needs to run, but it is constrained by the general sandboxing policies for browser extensions.

This threat model is different from the model commonly assumed by the browser vendors that browser extensions should always be "trusted" as long as they pass checks in a vetting process. We justify our threat model as follows. First, we should point out that the vetting process is never bullet-proof, which is acknowledged by browser vendors. For example, researchers from Google Chrome published their experience and lessons from years of fighting malicious extensions [1]. They reported that by using advanced detection techniques for the vetting process, the best detection rate that their system was able to achieve is 96.5%. Second, we argue that a dedicated study on malicious extensions should be useful. Many users are not aware that a malicious extension can cause harm to their security and privacy. On the other hand, some security researchers might think of the other extreme: once a malicious extension is installed, the security is all lost. In fact, malicious extensions are constrained by the sandboxing policies of the browser. The exact threats that a malicious extension can impose on users vary between the browsers, the underlying extension architectures and operating systems. The main aim in this paper is to establish a comprehensive and up-to-date understanding on precisely what malicious extensions can and cannot do in modern browsers.

The threats we present in our model are the expressed potential for the concurrence of a harmful event that can breach confidentiality, integrity and availability. To illustrate the threats, we implement a wide series of attacks which a malicious botmaster can launch by abusing the powerful APIs offered by the extensions. During the experiments, we also discover several inherent vulnerabilities, which make some browsers (e.g., Firefox) more susceptible to an attack than others (e.g., Chrome), which we will detail in Section 5.

## 4 Botnet-Based Attack Framework

Botnets are prevalent amongst the criminal underworld of the Internet. They serve as a tool to facilitate attacks against numerous victims at the same time. Historically, botnets made use of the Internet Relay Chat (IRC) protocol to distribute commands to the compromised zombie computers. However, with IRC falling into gradual disuse, the medium in which a botnet receives its commands is changing. Some solutions use simple transport layer protocols such as TCP or UDP, however, due to its simplicity and ease of access at the application layer, HTTP has become more popular as the next transmission medium [14].

### 4.1 The Framework's Architecture

We designed an architecture for complete control over a botnet of extensions, utilizing a simple Command & Control model (C&C) [15]. We illustrate the effect of our attacks using this architecture on Chrome, Firefox and Firefox for Android within our model. We illustrate in Figure 1 a component diagram representing an overview of our botnet system. Our framework provides the following functionality:

- An API schema which emphasizes independence from a target extension system, using inspiration from REST;
- Extensible command issuing system, along with data categorization utilities;

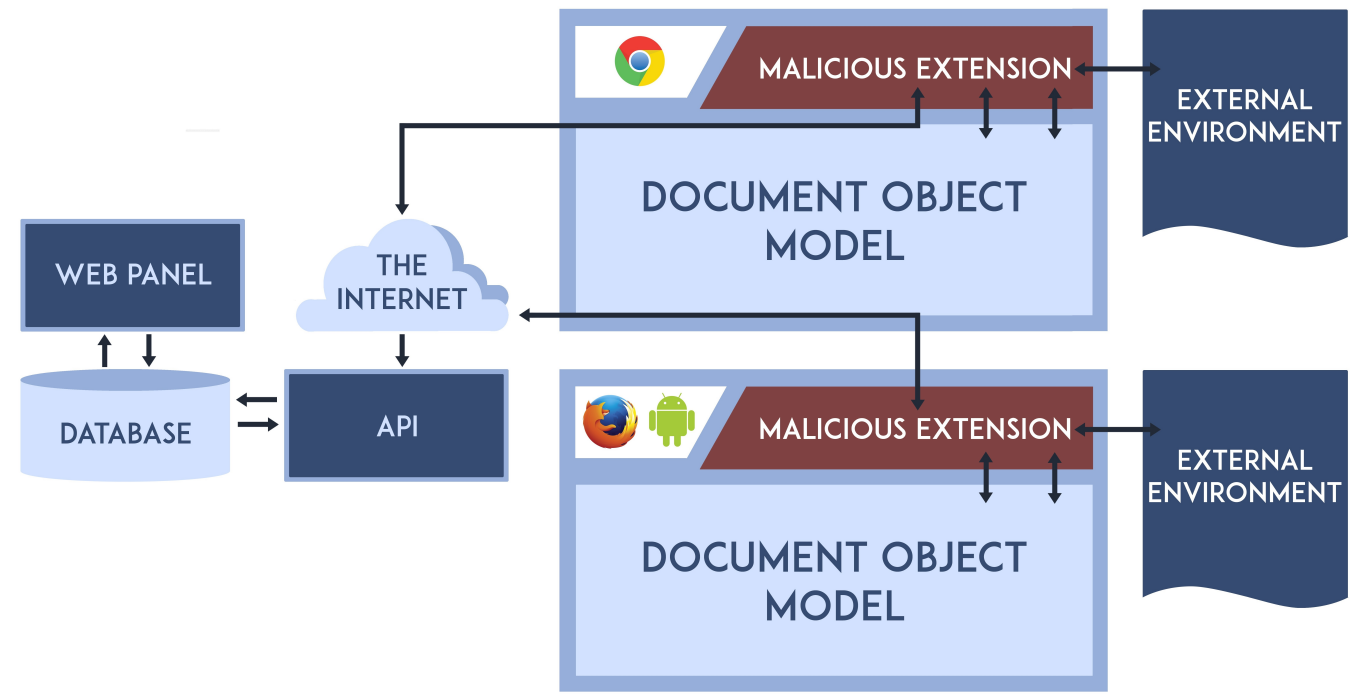

Fig. 1: Our botnet-based framework.

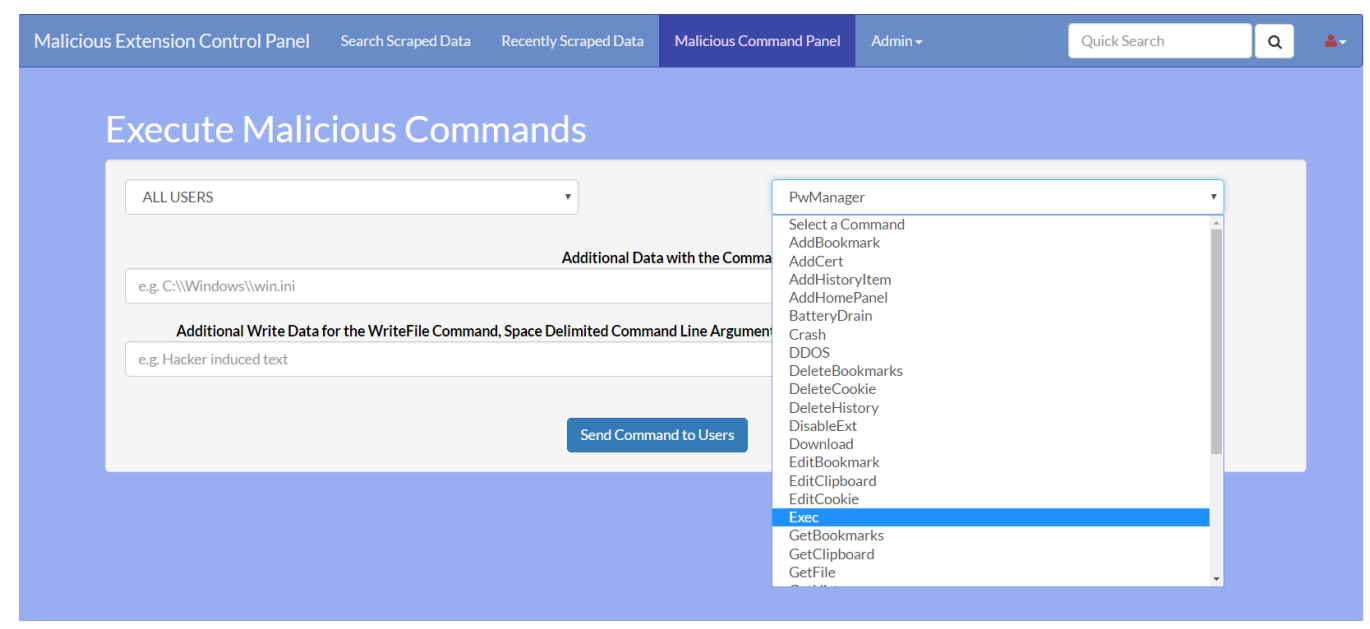

Fig. 2: The web panel for the bot master

- An easy to use graphic user interface (see Figure 2);
- Three examples of malicious extensions which exhaustively implement malicious capabilities respective to Chrome, Firefox and Firefox for Android.

### 4.2 The Framework in Operation

Each extension uses a *phone-home* beacon periodically to identify itself to the bot master's server. In our implementation we randomly generate and store a unique identifier that corresponds to one instance of an extension. If the user uses social media (say Facebook), we further tie the victim's identity to their social media account. We note that several other methods exist in identifying a user, including via IP address or fingerprinting techniques using metadata available to the browser. One advantage of our botnet framework is to allow automated testing on an exhaustive list of capabilities that a malicious extension may exploit. On receiving an identity, the botmaster's server will automatically find a list of commands associated with the identity and return it to the extension. The extension will automatically execute said command, and mark it as executed. Victim data is then sent to the botmaster, though we note that certain attacks operate passively. For example, HTML data is periodically sent to the remote server at a set interval.

### 4.3 Botnet Structures

We will briefly discuss in this section other common botnet paradigms, and how feasibly they can be implemented using extensions.

**Rallying Techniques.** Rallying refers to the process in which a C&C server discovers the domain/IP address of the central command server [15]. Hard-coding the domain name is the easiest solution, however obfuscation techniques using minifiers are a viable alternative when using extensions. Other masking techniques include using Dynamic DNS services. Using this method, an attacker can mask the real C&C address behind a series of DNS servers, in which the path to the command server can be made resilient to mitigate against law enforcement shutting a server down. The use of Domain Generation Algorithms can be deployed by including the generation algorithm in the source code of the extension.

**Peer-to-Peer Botnets.** Peer-to-Peer (P2P) botnets are a phenomena that developed after the popularization of P2P networks. Each node acts as both a client, and a server, distributing commands to each peer it has registered, as well as executing them. Browser-based P2P botnets are made possible by a new technology called WebRTC, which provides APIs for P2P data interchange. WebRTC allows web browsers to request resources not only from backend servers, but also directly from browsers of other users. Our experiments show that P2P botnets could be implemented using WebRTC on both browsers, which will make it harder for law enforcements to take down a botnet.

## 5 Threats Caused by Malicious Extensions

Within this section, we systematically categorize, describe and tabulate threats caused by malicious extensions, with an analysis against the CIA (confidentiality, integrity, availability) triad model. To illustrate the threats, we implement concrete attacks and verify them against Chrome, Firefox and Firefox for Android browsers on multiple operating systems including Windows, Linux and Android. Details of the attacks are presented below. See Table 1 for a summary of the results.

### 5.1 DOM Based Capabilities

The DOM has been highlighted in a variety of prior studies including [6], [7], [8], [16], [17] as a common source for malicious exploitation. Similarly, the near-unrestricted access to JavaScript functionality has also been discussed at length in previous studies as well as [2], [18]. Generally, these studies take the approach of exploiting *benign-but-buggy* extensions, rather than assuming the role of an attacker who has gained control over the extension's source code.

**Full DOM Access.** We confirm the prevalent belief that extensions currently offer minimal protection against full DOM manipulation. The case is slightly mitigated by Chrome's permission systems restricting content scripts to executing on sites declared within the manifest file. However, this is impinged on by extension developers setting broad permissive tags which allow content script execution on any webpage.

**Iframe Based Phishing.** Prior works have identified iframes as troublesome in making indirect cross-site requests [7] and a prominent malvertisting source [19]. However, we specifically observe their power in launching phishing attacks against HTTPS sites. Commonly, a user is instructed to check the URL of a website, as well as the information of the green lock next to the URL bar to verify that the integrity of the site has not been compromised and no intruder is launching a man-in-the-middle attack. Our iframe attack was able to seamlessly substitute the page with an attacker's phishing site, whilst giving the victim the credence that they were still browsing their original site. This was done without tampering either the URL bar, or the information shown within the lock dialog. We motivate this attack with an example, shown in Figure 3, illustrating the Facebook page substituted with the phishing page: *Phishbook*.

### 5.2 JavaScript Based Capabilities

JavaScript brings a plethora of functionality that can be of interest to an attacker. Listeners for keystrokes, mouse-strokes and touchstrokes were found to be applicable in attacks against both browsers. Furthermore, attacks against the availability of the browser via executing resource exhausting scripts were found to be effective, especially on Firefox, which has no recovery safeguard in place due to its monolithic process architecture. Furthermore, sensor based APIs such as HTML5's *geolocation* were requested as part of the website rather than the extension: thus if a user accepts access to their GPS device, the extension can read their geolocation coordinates.

**Unrestricted Cross-Site Requests.** We validate that the extension models we evaluated do not restrict cross-site requests made via XHR (XML HTTP Requests). XHR requests are the most powerful tool for a botnet's arsenal. Remote connections can be made to an attacker-controlled server for harvested data to be transmitted back, as well as for commands to be distributed across the botnet.

### 5.3 Cross-Site Scripting (XSS) vs. Extensions

XSS can be used to launch the attacks we have described in the previous sections as they are JavaScript based. An XSS attacker relies on finding a vulnerability within a site, where they can launch an XSS payload. An attacker using an extension can always launch the attack via the content script.

### 5.4 Cookie Capabilities

To mitigate the impact of XSS attacks, flags demarcating the cookie as being inaccessible by JavaScript are used, known as *HTTPOnly* flags. We find that extensions have access to more powerful cookie APIs which allow access to cookies even if they are marked as *HTTPOnly*, *Session* or *Secure*. Browsers allow access by extensions to these extended features as extensions are seen as "trusted" once installed, whereas regular JavaScript is not. By periodically transmitting cookie values back to the attacker's remote server, session hijacking attacks may be performed with relative ease.

### 5.5 Clipboard, Bookmark and History Capabilities

We validate the findings performed in [7] on Firefox, and apply the analysis on Chrome and Fennec. It was found that in Fennec, Mozilla have not yet implemented any Add-On SDK method of accessing the clipboard, bookmark or history, though a bypass for the clipboard was found

TABLE 1: Summary of capabilities which a malicious extension likely exploits to compromise a user's security and privacy.

| Capabilities Category (Botnet Commands) | Chrome (54) | Firefox (49) | Firefox Android (49) | CIA Triad Impact | Comments |
|---|---|---|---|---|---|
| **DOM-Based Capabilities** | | | | | |
| Read Web Page's DOM [6], [7], [16] | ✓ | ✓ | ✓ | Confidentiality. | Full read access to a page's DOM is possible. |
| Edit Web Page's DOM [6], [7], [16] | ✓ | ✓ | ✓ | Integrity, Availability. | Editing existing DOM elements is possible. |
| Write to Web Page's DOM [6], [7], [16] | ✓ | ✓ | ✓ | Integrity, Availability. | Appending new content to the DOM is possible. |
| Replace Web Page's DOM [6], [7], [16] | ✓ | ✓ | ✓ | Integrity, Availability. | Replacing an element of the DOM with another is possible. |
| Iframe Based Phishing | ✓ | ✓ | ✓ | Integrity, Availability. | Changing a page's entire DOM structure to an external iframe, without modifying the URL, is possible for both http and https sites. |
| **JavaScript-Based Capabilities** | | | | | |
| Crash Browser | Partial | ✓ | ✓ | Availability. | Chrome's multi-process system mitigates the attack to only the active tab. Firefox is subject to an entire browser crash. |
| Use of Eval [2], [7], [16], [18], [20] | Partial | Partial | Partial | Integrity | Chrome sandboxes instances of eval, though it can be disabled by editing the Content-Security Policy. Firefox's review process adopts a harsh extension rejection mechanism for extensions using eval. Eval can be abused to remotely execute tailored JavaScript code. |
| XHR Requests [6], [7], [16], [18], [21] | ✓ | ✓ | ✓ | Confidentiality. | XHR Requests can be used to transmit DOM content, or supply botnet commands. |
| Location Data [17] | ✓ | ✓ | ✓ | Confidentiality. | The request is made using JavaScript originating from a site, rather than the extension, so it is not unreasonable to assume that a user may click allow allowing for GPS coordinates to be scraped - if the site has originally requested it, then the decision is stored and is not requested again by the extension. |
| Keystrokes [7] | ✓ | ✓ | ✓ | Confidentiality. | Keystrokes can be captured and transmitted with relative ease. |
| Mousestrokes / Touchstrokes [17] | ✓ | ✓ | ✓ | Confidentiality. | Mousestrokes or touchstrokes on Android can impact on a user's privacy such as capturing websites which use graphical passwords. |
| **Cookie Capabilities** | | | | | |
| Read Cookies [7], [16] | ✓ | ✓ | ✓ | Confidentiality. | A full listing of cookies can be read, including HTTP-Only and Secure cookies. |
| Edit Cookies [7], [16] | ✓ | ✓ | ✓ | Integrity, Availability. | Cookies can be edited indiscriminately. |
| Delete Cookies [7], [16] | ✓ | ✓ | ✓ | Integrity, Availability. | Cookies can be deleted indiscriminately. |
| **Clipboard Capabilities** | | | | | |
| Read Clipboard | ✓ | ✓ | ✓ | Confidentiality. | Clipboard data can be read in both Chrome and Firefox. |
| Modify Clipboard | ✓ | ✓ | ✓ | Integrity. | Clipboard can be modified via the use of JavaScript in all browsers. |
| **Bookmark Capabilities** | | | | | |
| Read Bookmarks [7] | ✓ | ✓ | ✗ | Confidentiality. | A full listing of user bookmarks can be accessed. Fennec provides no native method to interact with bookmarks. |
| Add Bookmarks [7] | ✓ | ✓ | ✗ | Integrity. | Adding bookmarks, including folders, is possible, except in Fennec. |
| Edit Boookmarks [7] | ✓ | ✓ | ✗ | Integrity, Availability. | Current bookmarks can be traversed and updated, leading to malicious URL changes, except in Fennec. |
| Delete Boookmarks [7] | ✓ | ✓ | ✗ | Availability. | Chrome imposes restrictions on removing and adding bookmarks in the root folder, however, the rest are deletable. Not possible on Fennec. |

| Capabilities Category (Botnet Commands) | Chrome (54) | Firefox (49) | Firefox Android (49) | CIA Triad Impact | Comments |
|---|---|---|---|---|---|
| **Browsing History Capabilities** | | | | | |
| Read History [7] | ✓ | ✓ | ✗ | Confidentiality | Access to a user's browsing history is possible, though not supported yet in Add-on SDK for Fennec. |
| Write to History | ✓ | ✗ | ✗ | Integrity. | Firefox does not natively support writing entries to the browser history. |
| Delete History | ✓ | ✗ | ✗ | Availability. | Firefox does not natively support deleting entries from the browser history. |
| **File System Capabilities** | | | | | |
| Directory Listing [7], [16], [17] | ✗ | ✓ | ✓ | Confidentiality. | Directory listing is available on Firefox; Android's file system restrictions prevent some directories from being listed. |
| Read Files [7], [16], [17] | ✗ | ✓ | ✓ | Confidentiality. | Reading files, including system files, is available, though Android's default root permissions apply. |
| Edit Files [7], [16], [17] | ✗ | ✓ | ✓ | Integrity. | If the user permission in which Firefox was started with matches the file's access permissions, then editing is possible. |
| Delete Files [7], [16], [17] | ✗ | ✓ | ✓ | Availability. | If the user permission in which Firefox was started with matches the file's access permissions, then deleting content is possible. |
| Add new Folders [7], [16], [17] | Partial | ✓ | ✓ | Integrity. | Permission withstanding, adding folders to the file system is possible in Firefox. In Chrome, only subfolders may be added to the already configured downloads directory. |
| Add new Files [7], [16], [17] | Partial | ✓ | ✓ | Integrity. | Permission withstanding, adding files to the file system is possible in Firefox. In Chrome, downloading files programatically is plausible, but only to the configured downloads folder. Prompts are displayed on dangerous file downloads within Chrome. |
| Execute Processes [7], [16], [17] | ✗ | ✓ | Partial | Confidentiality, Integrity, Availability. | Executing files, including terminal commands is possible in Firefox. Android's scope is limited to only executing APK files. |
| **Extension Management Capabilities** | | | | | |
| Disable Extensions [7] | ✓ | ✗ | ✗ | Availability. | Can be done on Chrome silently, functionality was recently removed from Firefox. |
| Uninstall Extensions [7] | Partial | ✓ | ✓ | Availability. | Chrome requires user confirmation, whereas Firefox can uninstall extensions silently. |
| **Other Capabilities** | | | | | |
| Proxy Settings | ✓ | ✓ | ✓ | Confidentiality, Integrity, Availability. | Browser proxy settings can be modified to point to an attacker-controlled proxy. |
| Browser Preferences [7] | ✗ | ✓ | ✓ | Integrity. | Firefox exposes methods to change internal browser preferences. |
| DDoS [6], [17], [21] | ✓ | ✓ | ✓ | Availability. | XHR Requests are not rate limited thus extensions can be used for flooding-based DDoS attacks. Depends on JavaScript capabilities. |
| Password Manager [7], [16], [17] | ✗ | ✓ | ✓ | Confidentiality | If a user has not set a master password, plaintext passwords can be gathered silently. If the user has set a master password, then it will require entering it within the prompt. |
| XPCOM Usage [7], [11], [17], [22] | ✗ | ✓ | ✓ | Confidentiality, Integrity, Availability. | Can still be used in Jetpack extensions, though Firefox will be deprecating XPCOM soon. |
| System Library / API Usage [7], [11], [17], [22] | ✗ | ✓ | ✓ | Confidentiality, Integrity, Availability. | Low-level APIs such as WinAPI, Linux OS libraries, and the JNI on Android can be bridged and accessed by Firefox extensions. |
| Battery Drain [17] | ✗ | ✗ | ✓ | Availability. | It can be made so that a phone will not automatically go to sleep when left unattended. |
| Certificate Exceptions | ✗ | ✓ | ✓ | Confidentiality, Integrity. | An attack we introduce, and describe within the article. Has a dependency on file system capabilities. |

on the Android system utilising XPCOM functions. On Chrome and Firefox, it was found that full access to a user's browsing history was possible. Modifying or deleting items from the history was a privilege granted only to Chrome, though it is possible to use file capabilities to emulate this attack on Firefox by modifying the *places.sqlite* file directly. Bookmarks were found to be creatable, readable, updatable and deletable on both desktop browsers. The clipboard was found to be accessible and modifiable on all browsers. All three functions discussed in this section can be used to invade a target's privacy, as well as inject content such as illicit material into the target's browser.

## 5.6 Firefox Exclusive Attacks

From the literature analyzed in combination with our own analysis of Firefox and Fennec, we observe that its less restrictive permission model allows us to access the external environment of the operating system itself, as Firefox provides access to it via a series of APIs. The original development philosophy of Mozilla extensions was to make their extensions as powerful as the browser itself, and though over a decade of security analysis has been performed on the Firefox system which resulted in more restrictions being applied, we present some critical observations we have made within this section.

### *5.6.1 File System Capabilities.*

Firefox's arbitrary file system access has been a longstanding concern amongst the extension community. We confirm that Firefox still allows extensions to access the file system, resulting in the possibility of arbitrary file reading, writing or deleting. Modifying folders and manipulating existing files and directory data was also confirmed as possible, as well as executing files and commands. As there exists no data transmission barriers between reading a file, and an XHR request, an attacker can read the contents of a file and transmit them back using the botnet framework we have introduced. Another scenario an attacker could launch using our framework involves writing a file to a directory, then launching a command to execute the said file.

It is important to note that these capabilities operate on the premise of privilege inheritance. Hence, if the browser is started by a user that does not have access to writing to a certain directory, then the operation is refused. Firefox for Android is most strongly affected by this situation, as the Android operating system restricts a large portion of the file system from being read, written or executed for phones which are not rooted. A similar case can be argued for desktop operating systems, however, traditional operating systems like Windows allow less restrictions with executing and reading files, by default.

### *5.6.2 XPCOM and System Library Usage.*

XPCOM is a powerful native interface used within Firefox to provide endpoints for use by extensions, which allows access to native browser methods that are used in the browser's programming interface. These features remain exposed within Add-on SDK. We also make the observation that operating system libraries such as the WinAPI on Windows, or the JNI glue library on Android can be accessed with little restriction. This can result in an extension calling low level routines exposed in the operating system user space, such as accessing process memory.

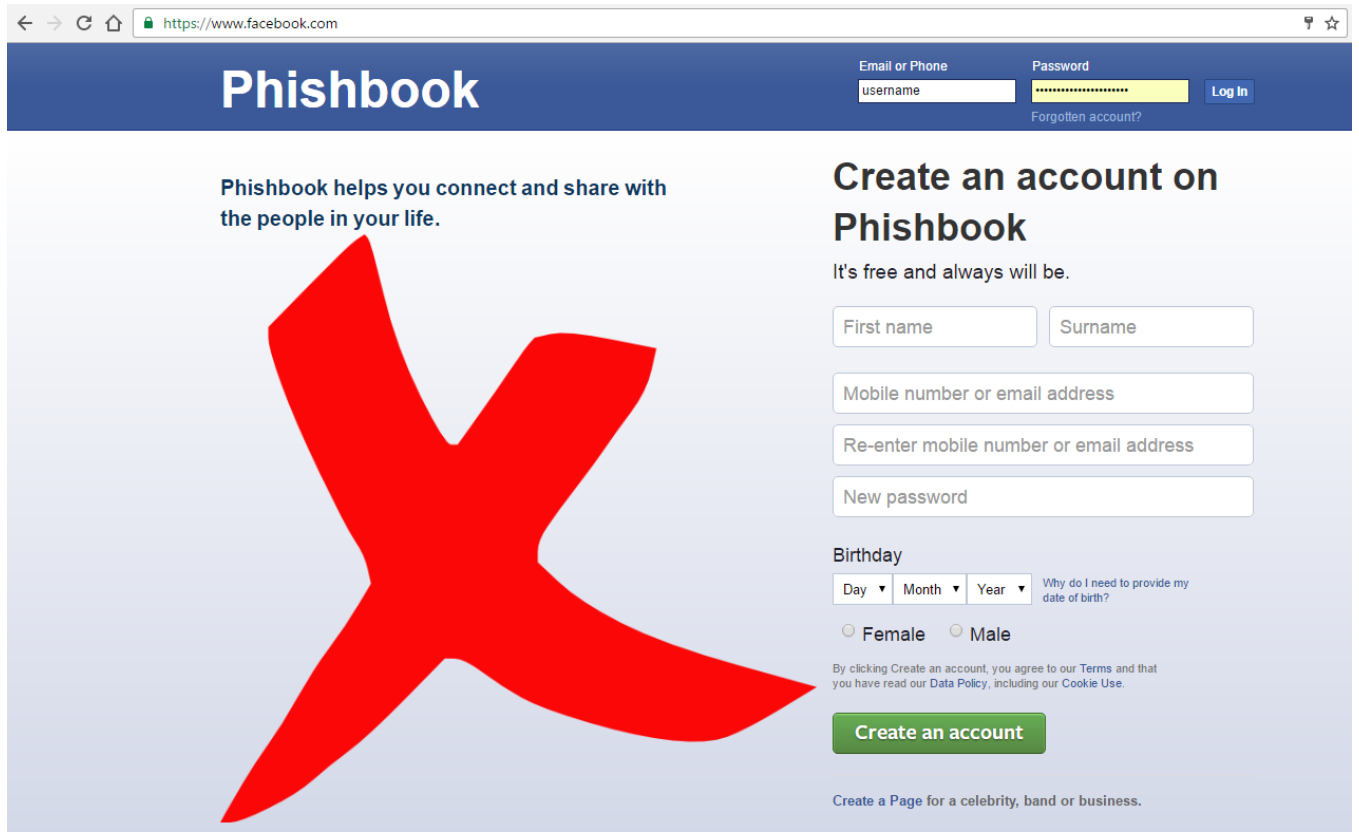

Fig. 3: The result of an extension substituting Facebook's page with "Phishbook" in an iframe based phishing attack. The browser continues to identify the page as Facebook (note the genuine *https://www.facebook.com* in the URL).

### *5.6.3 A Note on Firefox for Android*

Fennec's Add-on SDK port was generally designed to be backwards compatible with desktop Firefox. We thus observed a vast similarity in the range of attacks that could be performed on Android users who use Firefox. There exists one main study done on the security of Firefox for Android conducted by Marston et al. [17] in 2014. The researchers identified certain malicious capabilities within the Add-On system, as well as some predicted vulnerabilities in the future. We confirmed Marston et al.'s section of the study which focused on malicious extensions, and found that their results could be repeated. We also found that their predicted vulnerabilities, based on our experiments, remain currently not possible.

After applying Marston et al.'s study to our botnet framework, we applied the analysis we performed on Chrome and Firefox, to Fennec, and programmed the malicious functionality within a sample extension. We found, apart from the differences we have highlighted in Table 1 and the prior sections, that attacks using sensors were more plausible as mobile phones are far more likely to contain devices such as GPS in their hardware than a desktop computer. Furthermore, we also conducted an attack against battery drainage, by preventing the phone from auto-locking, which shows that performing denial of service attacks against power resources is also possible.

## 5.7 Certificate Exception Attack

One benefit of a systematic categorization of capabilities is that it enables us to discover how seemingly innocuous APIs can interact with the underlying system in an unexpected way to compromise the user's security. As a demonstration of this, we present a certification exception attack that has not been reported before.

The security of our online transactions critically depends on the HTTPS protocol and on the management of the public key certificates. A public key certificate is issued by a trusted certificate authority (CA) and cryptographically binds the site's identity with other details such as an expiry date in a digital signature. Within Firefox, including Fennec, it is possible for a user to manually specify a certificate exception via the user interface provided, if a site uses an untrusted or erroneous certificate.

Our proposed attack removes this element of interaction from the victim, and allows certificate exceptions to be added silently without the victim being privy to this knowledge. This is because Firefox insecurely stores certificate exceptions in a plain-text file located in the user profile in a file named *cert_override.txt*. An attacker can manipulate this file so that new certificate exceptions could be added. This attack was reproducible on Firefox for Android. We emphasize that this vulnerability does not exist in Chrome where certificate exceptions are stored on a per session basis.

As with our previous attacks, we integrated this attack so that it was possible to specify a site and certificate to add, remotely, via the botnet framework. It was found that expired and untrusted (such as self-signed) certificates, along with certificates containing URL mismatches could be silently marked as trusted using this attack.

This attack has several implications. Ordinary web users are often educated to check the padlock as a security sign of a web site. The browser must warn the user if there is any abnormality in the certificate. However, our attack shows that the presence of a padlock has little meaning when the exception warnings can be silently suppressed by an extension. Furthermore, if a certificate exception is added without the user's knowledge, then the attack can be leveraged to perform man-in-the-middle attacks [23]. Other implications include marking outdated certificates as valid, as well as allowing sub-domain mismatches within certificates to occur, which can be a problem in shared domains.

### 5.8 Extension Management Capabilities

First analyzed by Wang et al. [7] on Firefox, browsers provide extension mechanisms to manipulate and access data regarding other extensions present on the browser. Capabilities that allow an extension to disable or uninstall are the most pertinent to an attacker, as this could lead to the removal of anti-malware extensions from the victim's computer. It was found through our experiments that Chrome did not allow silent uninstallation of other extensions, opting to show a dialog of confirmation instead. However, it did allow for extensions to be disabled silently without notifying the victim. On Firefox, a reverse of this situation was observed. Disabling an extension was no longer possible; however, uninstalling extensions was found to be achievable. This situation applied to Fennec as well.

### 5.9 Other Capabilities

As we highlight in Table 1, a variety of other problematic APIs which a malicious attacker may abuse are identified. The proxy system was found to be editable for protocols such as HTTP and HTTPS on all three browsers examined. This means that traffic could be routed to an attacker-controlled server and examined. Attacks against the availability of a user's Internet are also possible as an attacker can change the proxy address to an inexistent IP address, causing connections to fail.

Firefox, including Fennec, exposes an API to the password manager system. These features were identified by Wang et al. [7], and we verify them as still currently functional. By default, the password manager does not encrypt the password store with a master password until the user sets one up; this results in unrestricted access to passwords stored in the manager. If a user has set up a password manager, then the attack is still effective, as long as the victim enters the password within the prompt offered by Firefox and the extension will subsequently gain access to all the stored passwords.

### 5.10 Bypassing the Web Stores

We managed to upload a customized extension to Chrome's web store using the update system. The extension has the iframe-based phishing attack, which reads the command from our web server. We uploaded a similar extension to Mozilla's add-on store, based on the certificate exception attack. This extension also was designed so it could read commands from our web server. During the review in the vetting process, we disconnected the server so that no one would be affected by our (proof-of-concept) attacks. We removed both extensions as soon as they passed the vetting process before anyone else was able to download them, following the guideline from our university ethics committee.

To bypass the vetting process, we masked the malicious extension as a benign extension via the use of clean, structured and well commented code that indicated the extension served a valid research purpose. Although in the comments, we documented that the extension was written for a research project, we wanted to test if the extension could pass the automated checks by Chrome and Firefox. The vetting process by Chrome is primarily automated, and the extension we submitted there was quickly approved. Firefox has an additional stage of manual review, which took longer. But in the end, the extension was also approved.

We disclosed our findings to Google and Mozilla. Google replied by highlighting their web store policy regarding that they remove malicious extensions. However, our experiments indicated that uploading a malicious extension controlled by the botnet was possible. As of the time of writing, Mozilla has yet to reply.

## 6 Comparison with Malware

Malware is a term that aggregates and identifies programs which exhibit malicious behaviors towards their victims. As compared with traditional malware, an extension-based botnet has the advantage that administrator privileges are not required for installing the extension. Furthermore, once installed, extensions effectively form an integral part of the browser and hence can easily avoid detection by anti-virus software. In this section, we will compare our extension-based botnet with other malware according to the taxonomy of malware behaviors presented by Gregio et al. [24].

**Viruses, Worms and Trojans.** Viruses can be produced via extensions, though are more effective on Firefox due to its capability to write and execute files. Traditional worms face some restrictions in propagating via extensions. A classical worm will self-propagate, typically via a known exploit that requires little to no user interaction to spread the worm. To successfully launch a browser worm, exploits would need to be found within the browser's core logic. With this in mind, an adversary may still distribute worms via Firefox's extension system, due to its control of low level APIs. To give a contemporary example, the ransomware WannaCry exploited a Windows flaw [25], and then spread itself automatically using worm-like behavior by abusing the flaw in unpatched systems. With Firefox's capabilities to make low-level API calls, it's not too hard to envision the possibility of loading worms like WannaCry into an extension. However, this method would not be currently possible on Chrome. Finally, trojans can be created with relative ease and such was the case documented by Utakrit [26] in his analysis of extension-based banking trojans.

**Botnets and DDoS.** We refer the reader to Section 4 for a detailed analysis of botnets, and different architectures. Botnets are often used to launch Distributed Denial of Service (DDos) attacks. To assess the feasibility, we have implement a DDoS attack based on the use of XHR as in [21] for both extension systems. We observed that an average of 252 XHR requests could be sent per second, from one machine. With some established botnets having reached over 180,000 [27], this number could be amplified to a maximum 45,360,000 requests per second. With such a number of requests, flooding-based DDoS becomes a possible attack that can be performed by extension botnets.

**Spyware, Adware & Ransomware.** Table 1 documented a large portion of features that can be used to break the user's confidentiality. For example, with liberal access to the DOM possible, Spyware can be written to simply transmit the user's browsing contents to the central command server on all three targeted browsers. Cookies, browsing history, bookmarks, keystrokes from within the browser focus, location data and passwords can all be captured by extension-based Spyware. Adware can be further introduced by breaking the integrity of the browser or DOM; within our implementation we provide a *Spam* attack, that enables the injection of an attacker-defined HTML ad into the browser's current page. Ransomware cannot be effectively created in Chrome extensions, as it could only encrypt the contents of a webpage. Within Firefox, ransomware could be created to encrypt some of the file system's contents leaving the victim unable to access their documents without paying a ransom for the decryption key.

**Rootkits.** Rootkits by their traditional definition give the attacker a way of acquiring root or system-level permissions as a result of an exploit. This type of malware would be unlikely to deploy using extensions, especially in Chrome due to its isolated worlds model and Fennec as Android typically blocks root access by default. Such an attack would require a major vulnerability within the browser. Another possibility is that a Firefox extension can download and write a rootkit to the file system and execute it. However it would still require the user to confirm the escalation of privileges dialog, which can raise the alert.

**Surreptitious Software.** Code can be obfuscated in extensions by using a JavaScript minifier, or remotely loading the extension code from an attacker-controlled source. The use of obfuscation is commonly used to bypass Chrome's extension review process, whereas Firefox's review process can be bypassed by writing code that appears benign and utilizes social engineering to mislead the reviewers of its benignity. In the extension we uploaded to the Firefox add-on store, we heavily made use of comments describing to the reviewers that the extension was for internal use within our university only.

## 7 Countermeasures

Having presented a thorough overview of the threats present in modern extensions, we now take a step back and discuss a range of countermeasures. Some of the threats/vulnerabilities identified in this paper can be easily addressed, but for many others, fundamental changes in the underlying extension architecture are needed, which requires further research.

**Restricting Arbitrary Access to the DOM.** An extension of the privilege management system on Chrome, was proposed by Liu et al.'s paper [6]. They introduce the concept of *sensitivity* as a default protection mechanism against DOM elements which typically contain confidential information, such as password inputs. Sensitivity levels are by default applied to HTML elements according to a default mapping between levels and elements. For example, high sensitivity may be applied to input tags containing the password attribute, whereas medium sensitivity is applied to tags containing IDs or usernames. By default, other elements are marked as having a low sensitivity. Content scripts are marked with a default sensitivity level. This level corresponds to the elements from which they may read a value. If a content script has lower privileges than the sensitivity of the element it is attempting to read, the read will be rejected.

**Iframe Based Phishing.** Using an extension to substitute or modify a page with an iframe that loads external content, especially on an HTTPS site, should be detected. This would add negligible overhead, as it would only require checking if an iframe is being inserted by a content script. For further usability, a user may be presented a confirmation dialog clearly illustrating the domain where the external content is loaded from which they may then chose to allow or disallow. As this would only apply to iframes inserted by extensions, this would not affect advertising or legitimate web-uses of iframes.

**Defeating the Botnet: Restricting Cross-Site Requests.** Cross-site requests are a dangerous tool as they allow sensitive information to be transmitted to a remote location. Proposals to curb the damage done by cross-site requests have been introduced by several studies. Liu et al.'s [6] aforementioned paper introduced a *micro-privilege* management schema which separates cross-site access between the extension core and the content scripts. The extension core is granted explicit cross-site access privileges via the manifest file, whereas any attempt to bypass this via the use of a malicious content script is prevented by having it explicitly declare any remote origins added to the web page. This schema informs the user which sites the extension has access to in transmitting information.

**Over-Permissive Browser APIs.** The main countermeasure in effect against the arbitrary use of JavaScript APIs is permission management. This method is only in use within Chrome, as it is necessary to explicitly declare the APIs which the extension uses, in the manifest file. Permission systems however are not a silver bullet to preventing the malicious use of APIs. Application permission systems experience multiple shortcomings, with adware developers using enticement techniques and look-alike naming schemes to push more invasive permissions out to users [12]. Furthermore, additional evidence exists that permissions become less effective over an installed application's lifetime.

We recommend some minor changes to Chrome's permission system by marking more dangerous permissions with greater prominence on the installation prompt, rather than presenting them equally as if they have analogous security implications. This is an idea introduced by Felt et al [13]. Ideally, no read data from browser APIs should be transmitted via a cross-site request, but enforcing this would not be practical by current technology. We therefore take a moderate stance in suggesting a combination of the countermeasures suggested in managing cross-site requests, alongside a well designed permission system. We suggest in particular that Mozilla should follow this schema within WebExtensions.

**Firefox and Fennec's Privileged Extension Model.** Currently, Mozilla are attempting to phase in WebExtensions, a modernization of the previous extension systems. From recent analysis, it appears Mozilla has removed APIs that allowed access to the file system and other contexts outside the browser. However, their success will rely on the extension developers to migrate to using the proposed WebExtensions kit. There are no dates currently on whether Jetpack will be fully removed or operate in parallel to WebExtensions.

**Firefox's Certificate Exception Attack.** Certificate exceptions should not be added by extensions, nor stored in plaintext. The *cert_override.txt* file poses a suitable target for malware running with non-superuser privileges. We recommend that the method of providing certificate exceptions should follow a system similar to that of Chrome, where certificate exceptions are always approved by the user, on the per session basis.

## 8 Related Work

In this section, we review the related work. Although the threats of malicious extensions have been reported before, we are not aware of any systematization of knowledge about effects of malicious extensions against major browsers across multiple operating systems. Our paper contributes to this subject by systematically analyzing existing literature, as well as running experiments enumerating an exhaustive range of APIs that can be abused for malicious purposes.

**Malicious Browser Extensions.** Prior work was conducted in this area on the Google Chrome browser by Liu et al. [6] who demonstrated the power that could be exploited from Chrome extensions as part of a rudimentary botnet. An empirical study of dangerous behaviours on Firefox extensions was performed by Wang et al. [7], in which a set of 2465 web store extensions were examined against an array of security-critical functionality of varying threat levels. Ter Louw et al. detail a comprehensive browser extension study in 2008 [22], though it largely focuses on the XPCOM architecture. We also note the works done by Acker et al. [28], who identify malicious and vulnerable scripts within the popular Firefox community scripting extension, Greasemonkey.

**Exploiting Benign Extensions.** Works were done by Barth et al. [16] in designing the extension model, that has been adopted by Chrome. Their paper proposes an adapted version of their research to Firefox. Carlini et al. [18] evaluated the Chrome extension architecture for features which would bypass or compromise the principles proposed by Barth et al. Following that analysis, they produced an evaluation of 100 randomly selected Chrome extensions to audit them for hidden vulnerabilities, and discovered at least 40% had at least one vulnerability which could be exploited.

**Extension Vulnerabilities.** Generally, vulnerability exploitation is reported via Chrome and Firefox's respective vulnerability programs. Publishing an article about the vulnerability is left at the discretion of the reporter. However, there is adequate academic research targeted towards exploiting vulnerabilities within extensions themselves. Liverani and Freeman [11] further demonstrate XSS attacks against Firefox's *chrome* privilege zone.

**Static and Dynamic Analysis Techniques.** Static analysis techniques were introduced by Bandhakavi et al. [20], creating the vetting tool *VEX*. The tool identifies potentially malicious flows in Firefox extensions by analysing the source code. Dynamic analysis tools were introduced by studies such as those conducted by Dhawan and Ganapathy [10] and Kapravelos et al. [2]. These tools are inbuilt into the browser and involve tagging and monitoring objects which have an untrusted source as its provenance. Both of these analysis techniques can be used to diminish attempts to perform malicious actions successfully, though it is unknown how serious the impact of code obfuscation is.

## 9 Conclusion

We have introduced a browser extension based botnet framework, which we implemented as a way of demonstrating the impact of running botnets in Chrome, Firefox and Firefox for Android Jetpack extensions. We enumerate the list of malicious capabilities of these extensions, and discuss countermeasures to the identified security problems.

## Ethics and Code Availability

All of our experiments were approved by the university ethics committee. The proof-of-concept attacks were performed on the authors' own computers and used freshly created facebook accounts only for the testing purpose. For obvious reasons, we cannot release the developed extensions as open-source code, but we intend to make the code available, upon request, to researchers working in this field.